\documentclass[referee]{cjaa}           

\usepackage{graphicx}                   
\input{epsf.sty}                        
\input{psfig.sty}                       

\begin{document}

   \title{GRBs-SN and  SGR-X-Pulsar as blazing   Jets
}

 \author{Daniele Fargion
      \inst{}\mailto{}
      }

   \offprints{D. Fargion}                   

   \institute{Physics Department, University of Rome P.le A. Moro,2, 00185 ROME,
   ITALY\\
             \email{daniele.fargion@roma1.infn.it }
             }


 %


\abstract{Old and recent puzzles of GRBs and SGRs find a solution
with a model based on the fast blazing of very collimated thin
gamma Jets. Damped oscillating afterglows in GRB$030329$ find a
natural explanation assuming a very thin Jet -
$\frac{\Delta\Omega}{\Omega}\leq 10^{-8}$ - whose persistent
activity and different angle of view maybe combined at once with
the Supernov\ae ~power and the apparent huge GRBs output:
$\dot{E}_{GRBs} \simeq$
$\dot{E}_{SN}$$\frac{\Omega}{\Delta\Omega}$. This  leads to a
better understanding of the remarkable GRB-Supernov\ae ~connection
discovered in the GRB$980425$/SN$1998$bw and in the most recent
GRB$030329$/SN$2003$dh events.  The same thin beaming offer an
understanding of the apparent SGR-Pulsar power connection:
$\dot{E}_{SGRs} \simeq$
$\dot{E}_{Xpuls}$$\frac{\Omega}{\Delta\Omega}$. A thin collimated
precessing  Gamma Jet model for both GRBs and SGRs, at their
different scaled luminosity ($10^{44} - 10^{38}$ $erg \cdot
s^{-1}$), explains the existence of few identical energy spectra
and time evolution of these  sources leading to a unified model.
Their similarity with the  huge precessing Jets in AGN, QSRs and
Radio-Galaxies inspires this smaller scale SGR-GRB model. The
spinning-precessing Jet explains the rare ($\approx 6\%$)
mysterious X-Ray precursors in GRBs and SGRs events. Any large
Gamma Jet off-axis beaming to the observer might lead to the
X-Flash events without any GRB signals, as the most recent
XRF$030723$. Its possible re-brightening would confirm the
evidence of the variable pointing of the jet in or off line
towards the observer.
  Indeed a  multi-precessing Jet at peak activity in all
bands may explain the puzzling X or optical re-brightening bumps
found in the  GRB $021004$, GRB$030329$ and the SGR $1900+ 14$ on
$27$ August $1998$ and once again on the $18$ April $2001$. Rarest
micro-quasars neutron star in our galaxy as SS433, and Herbig Haro
objects and Cir-X-1  describe these thin precessing Jet imprints in
the spectacular shapes of their relic nebulae.}
 \authorrunning{D. Fargion}            
   \titlerunning{Puzzling afterglow's oscillations in GRBs and SGRs:
 tails of precessing Jets}  


   \maketitle
%

\section{Introduction:  Fire-Ball Models implosion}           
\label{sect:intro}
The clear evidence of gamma polarization in the $\gamma$ signals
from GRB$021206$ (Coburn \& Boggs) proves in the eyes of most
skeptical Fireball theorists the  presence of a thin collimated
jet (opening angle $\Delta\theta \leq 0.6^o$;
$\frac{\Delta\Omega}{\Omega}\leq 2.5 \cdot 10^{-5} $) in Gamma
Ray Bursts (GRBs). The time and space coincidence
 of  GRB$030329$ with SN$2003$dh definitely confirms   the
 association of GRB and Supernovae discovered in the GRB$980425$/SN$1998$bw event.
Therefore the extreme GRBs  luminosity is just the consequence of a
collimated gamma Jet observed on axis during a Supernova event.
Nevertheless the maximal isotropic SN power, $\dot{E}_{SN}$$ \simeq
10^{45}$ erg $s^{-1}$, should be collimated even into a  thinner
jet $\frac{\Delta\Omega}{\Omega}\leq 10^{-8}$, in order to explain
at the same time the apparent observed maximal GRBs output,
$\dot{E}_{GRB}$$ \simeq 10^{53}$ erg $s^{-1}$. Consequently
one-shot thin Jet  GRBs needs many more events - $\dot{N}_{GRBs}
\simeq$ $\frac{\Omega}{\Delta\Omega}\geq 10^{8}$ - than any spread
isotropic Fireballs, a rate that even exceeds the one of the
observed Supernovae. To overcome this puzzle a  precessing  jet
with a life-time $\tau_{Jet}$ $ \geq 10^3$ $\tau_{GRB}$ is
compelling. Relic GRBs sources may be found in compact SNRs core,
as  NS or BH jets;  at later epochs, their "weakened" $\gamma$ jets
may be detectable only within galactic distances, as Soft Gamma
Repeaters (SGRs) or anomalous X-ray Pulsars, AXPs. This common
nature may explain some connections between GRBs and SGRs, given
that rare spectra of SGRs show similar properties  to  GRBs. Also
X-Ray precursors detected in  both  GRBs and  SGRs suggest the need
for a precessing Jet model. A surprising multi re-brightening
afterglows observed in early and late GRB $030329$ optical
transient, like in the $27$ August $1998$ and $18$ April $2001$ SGR
$1900+14$ events, might be the damped oscillatory imprint of such a
multi-precessing $\gamma$-X-Optical and Radio  Jet.

\section{The new GRBs puzzles}
\label{sect:the diff}
The Gamma Ray Burst mystery lays  in its huge energy fluence, sharp
variability, extreme cosmic distances and very different
morphology.  A huge isotropic explosion (the so called Fireball)
was the ruling  wisdom all along last decade.  However the observed
millisecond time scales called for small compact objects, so
confined to become opaque to their own intense luminosity (over the
Eddington limit) because of the abundant pair production, and so
small in size and masses (few solar masses) to be  unable to
produce the large isotropic energies needed. Moreover, the spectra,
 had to be nearly thermal in a Fireball, contrary to the data evidence.
The Fireball became an hybrid and complex model, where power law
after power law, it tried to fit each GRBs spectra and time
evolution. The huge GRBs power as in GRB$990123$, made the final
collapse of the  model. New families of Fireball models including
Jets collimated within a $10^o$ beam have been introduced (as
Hyper-Nova, Supra-Nova, Collapsar ), which allow to lower, because
of the beaming, the energy budget requested.
However the apparent required GRB power  is still huge ($10^{50}$
erg $s^{-1}$), nearly $10^5$ more intense than other known maximal
explosion events (such as the Supernova one). Further evidence in
the last few years  have shown that Supernova might harbour a
collimated Jet Gamma Ray Burst (GRB$980425$/SN$1998$bw
,GRB$030329$/SN$2003$dh). To combine the Super-Nova Luminosity and
the apparent huge GRBs power one needs a very much thinner beam,
as small as  $\Delta\Omega/\Omega \simeq$ $10^{-7}$ or $10^{-8}$
respect to $\Omega \simeq 4 \pi$, (corresponding to a Jet angle
$0.065^o-0.02^o$ ). There is a statistical need (Fargion 1999) to
increase the GRB rate inversely to the beam Jet solid angle. The
needed SN rate to explain GRBs may even exceed the observed one,
at least for SN type Ib and Ic  ($\dot{N_{NS}}$ $\leq 30 s^{-1}$).
Assuming that only a fraction of the SN  (at most $0.1$)
experiences an asymmetric Jet-SN explosion,
 the corresponding observed rates $\dot{N_{GRBs}}$ $\simeq 10^{-5}
s^{-1}$ and $\dot{N_{SN}}$ $\simeq 3 s^{-1}$  imply
$\frac{\dot{N_{GRBs}}}{\frac{\Delta\Omega}{\Omega}} \simeq 10^{2}
s^{-1} \longleftrightarrow 10^{3} s^{-1}$, which is  nearly $2-3$
orders of magnitude larger  than what is observed for SN events.
In this scenario one must assume a GRB  Jet with a decaying
life-time (to guarantee the energy conservation) much larger than
the observed duration of GRBs, at least $\tau_{Jet} \simeq
10^{3}\tau_{GRBs}$.

We considered GRBs (as well as Soft Gamma Repeaters SGR) as
originated by very thin ($\leq 0.1^o$) spinning and  precessing
Jets (Fargion 1994; Fargion \& Salis 1995a, 1995b; Fargion et al.
1996a; Fargion 1999; Fargion 2001; Frontera \& Hjort 2001). In
this scenario GRBs are born within a Super-Nova,  collimated in a
very thin beam which make them glow with an apparent GRB
intensity. The inner geometrical dynamics of the spinning and
precessing jet, may explain the wide $\gamma$ burst variability
observed in different events.

\begin{figure}\centering\includegraphics[width=8cm]{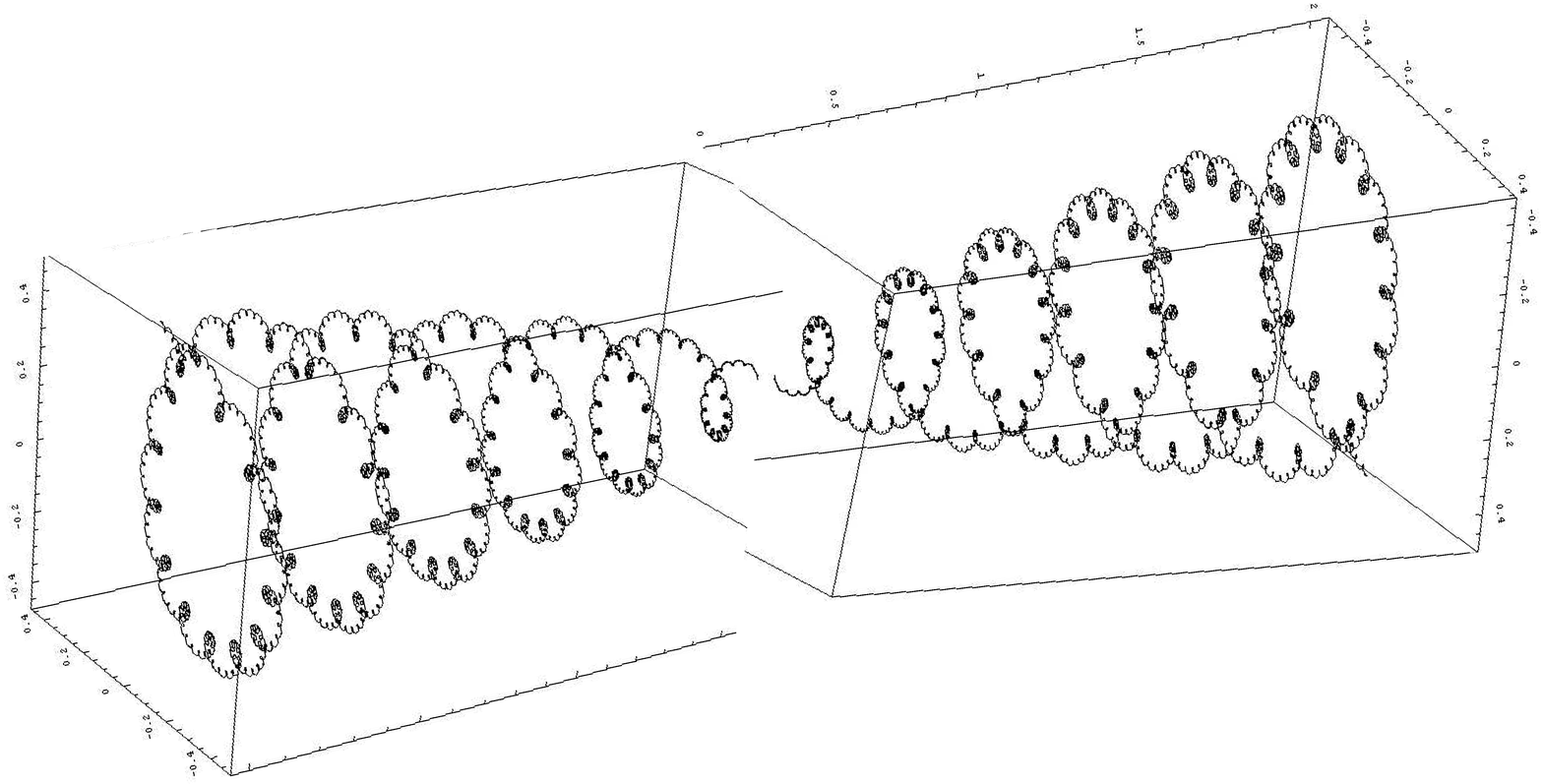}
\caption {A possible inner 3D structure of a multi precessing Jet.
The conical structures and the  stability at late stages  may be
reflected in the quasi-periodic repetitions of the  Soft Gamma
Repeaters whose emission is beamed   toward the observer. Its early
blast at maximal SN output may simulate a brief blazing GRBs event,
while a fast decay (in a few hours time-scale) may hide its
detection below the threshold, avoiding in general any GRB
repeater.} \label{fig:fig1}
\end{figure}

The averaged $\gamma$ jet deflection from the  line of sight
defines a power law decay; an inner damped oscillatory substructure
may be observed, as the peculiar damped oscillating afterglows in
GRB$030329$.  The thin, collimated ($\theta \leq 0.05^o$) and
long-lived jet (decaying in a few hours as a power law with index
$\alpha \simeq -1$)  spinning and precessing at different
time-scale, may better explain the wobbling of the GRBs and the
long sequence of damped oscillations of the X afterglows within
hours, and of the optical transient within days and weeks. The GRBs
re-brightening are no longer a mystery  in a one-shot model. These
wobbling signatures may be also be found in the rarest and most
powerful SGRs events. The spread and wide conical shape of these
precessing twin jets may be recognized in a few relic SNRs as in
the twin SN 1987A wide external  rings, the Vela arcs and the
spectacular Egg Nebula dynamical shape.
\begin{figure}\centering\includegraphics[width=8cm]{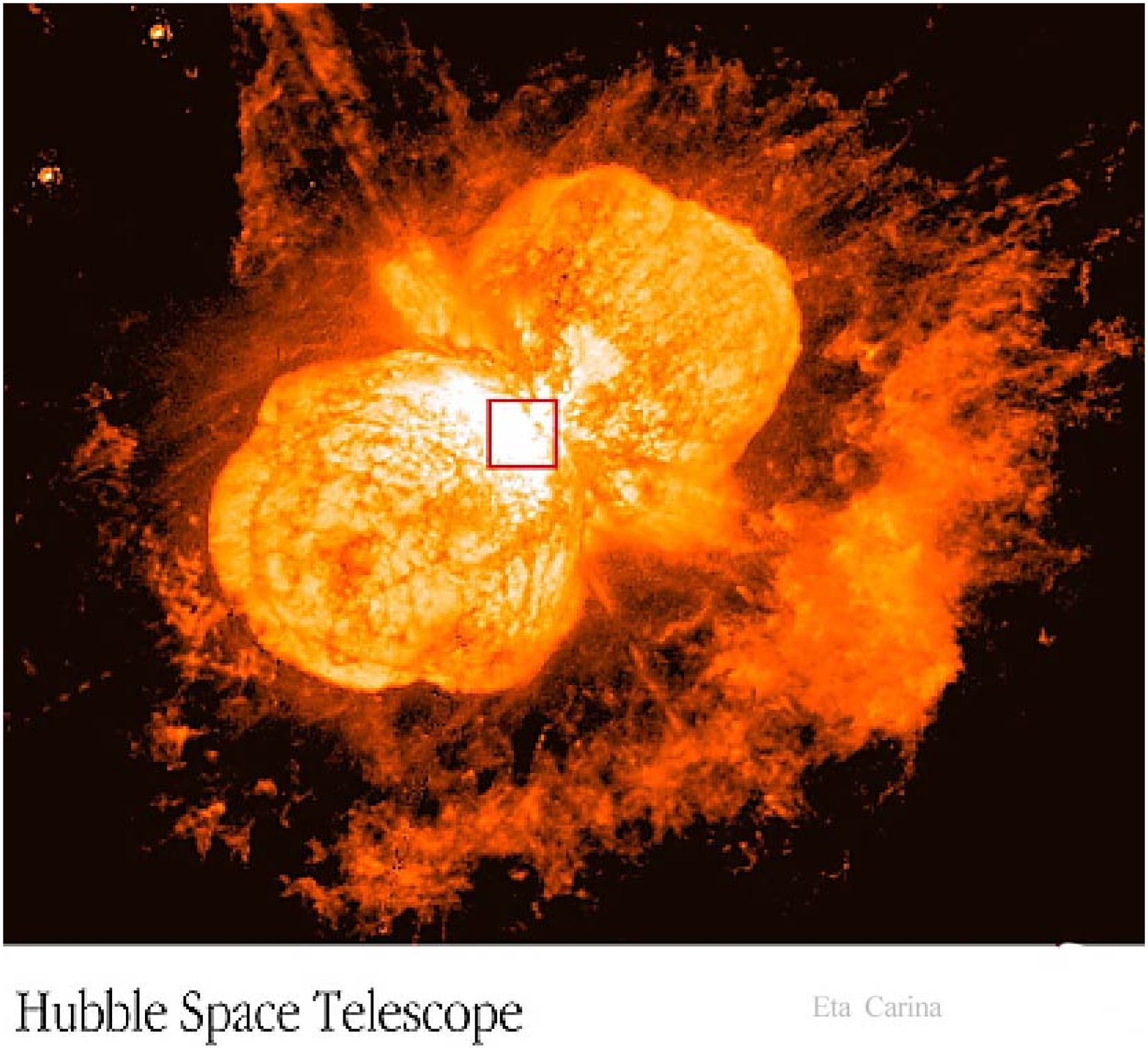}
\caption {Eta Carina Conical Hour Glass Shape. Such a mysterious
twin lobes may be  originated by an inner thin precessing Jet
hidden in the center of the nebula. The blow up of the nebula
could be due to the pressure of a  multi precessing jet  leading
to a twin Homunculus or a Hour-Glass  Nebula. The presence of the
thin jet escaping from the center has been observed at certain
angles of view (finger-like structures) (Redman, Meaburn,
Holloway 2002); the bending of the jet is due to an accretion
disk or to a compact binary companion.}
\end{figure}

\section{The  geometrical multi-precessing Gamma Jet in GRB}
\label{sect:the geom}
We imagine the GRB and SGR nature as the early and the late stages
of jets fueled by a SN event first and then by an asymmetric
accretion disk or by a companion star (white dwarf,WD, or neutron
star, NS). The ideal spinning Jet points  in a fixed direction;
however the presence of a companion star influences the stability
of the  jet. Indeed a   binary system angular velocity,
$\omega_b$, affects the beam direction  whose bending angular
evolution may be described in a  simple case as follows:
$\theta_1(t) = \sqrt{\theta_{1 m}^2 + (\omega_b t)^2}$. More
generally a multi-precessing angle $\theta_1(t)$ (Fargion et al.
1996a, 1996b) may be described by:
$\theta_1(t)=\sqrt{\theta_{x}^2+\theta_{y}^2 } $
$$
  \theta_{x}(t) =                               
  \theta_{b} sin(\omega_{b} t+ \varphi_{b} )+
  \theta_{psr}sin(\omega_{psr} t +  \varphi_{psr})+
  \theta_{N}sin(\omega_{N} t  + \varphi_{N})
$$
\begin{equation}
\label{eqn:eqn1}
  \theta_{y}(t) = \theta_{1 m}+
  \theta_{b} cos(\omega_{b} t + \varphi_{b})+
  \theta_{psr} cos(\omega_{psr} t +  \varphi_{psr})+
  \theta_{N} cos(\omega_{N} t  + \varphi_{N})
\end{equation}
where $1/\gamma $ is the characteristic jet opening angle, and
$\gamma$ is the Lorentz factor of the $\theta_{1 m}$ is the
minimal angular distance (impact parameter angle) of the jet
pointing toward the observer, $\theta_{b}$, $\theta_{psr}$, and
$\theta_{N}$ (all proportional to  $1/\gamma $)  are respectively
the maximal precessing angles due to the binary system,  the
spinning pulsar, and the nutation mode of the multi-precessing
axis of the jet. The arbitrary phases $ \varphi_{b}$, $
\varphi_{psr}$, $\varphi_{N}$ for the binary, spinning pulsar and
nutation,  are able to fit the complicated GRBs flux evolution in
most GRB event scenario. It is possible to increase the number of
the parameters with a fourth precession angular component whose
presence may better fit the wide range of variability observed.
Here we shall stick to a three parameter precessing beam.
\begin{figure}\centering\includegraphics[width=8cm]{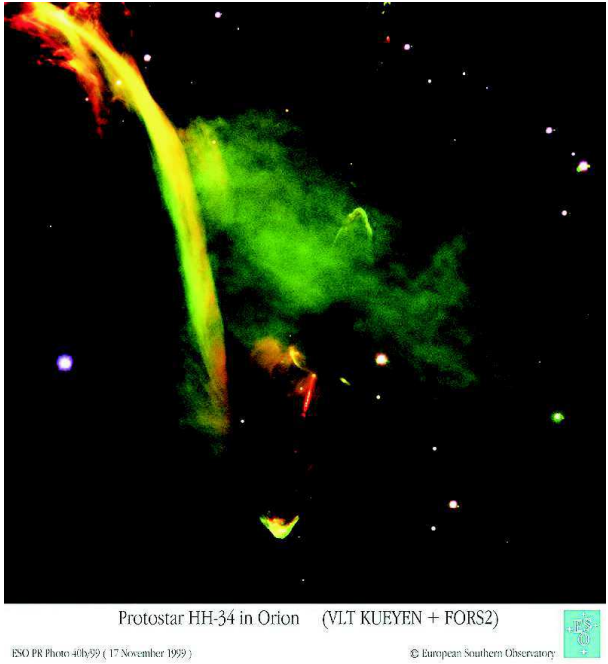}
\caption {The whip-like jet of HH34 micro-quasar. The long tail
visible both in the front and  behind the star describes a thin
moving jet. An internal  spinning sub-structure may be hidden
inside the width of the tail.}
\end{figure}

\begin{figure}\centering\includegraphics[width=8cm]{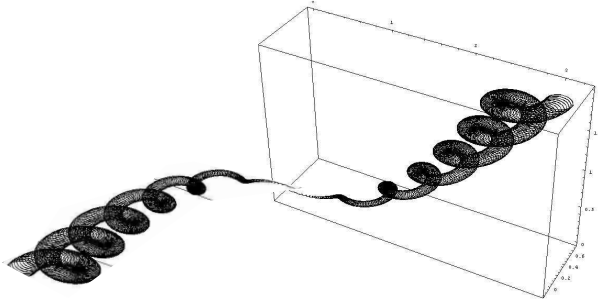}
\caption {A twin spinning and precessing jet  whose
 appearance along the line of sight,  at maximal supernova
output, produces the sudden GRBs. At later stages, the less
powerful jet would rather appear as a SGRs, visible only within
nearby, galactic-like distances}
\label{jet3D}
\end{figure}

\begin{figure}\centering\includegraphics[width=8cm]{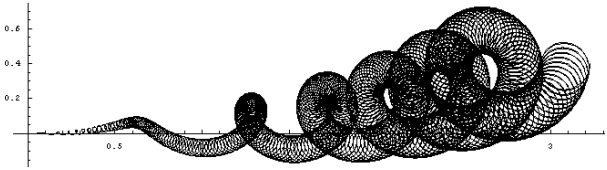}
\caption {The twin spinning and precessing jet configuration
projected onto a $2$dimensional screen.} \label{jet2D}
\end{figure}

\begin{figure}\centering\includegraphics[width=8cm]{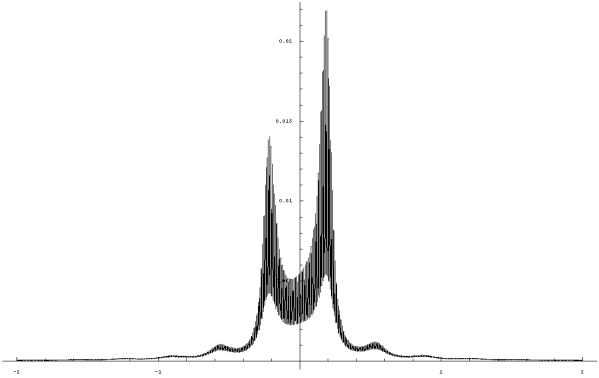} 
\caption{ Multi-bump afterglow behaviour of the precessing Jet
shown above, with the characteristic damped oscillatory decay as
found in the  GRB$030329$ and the  SGR of the $27$ August 1998. The
luminosity starting time is assumed near zero (at SN birth time).
In the present simulation the assumed Lorentz factor is
$\gamma_e$$= 2 \cdot 10^3$ }
\end{figure}
\begin{figure}\centering\includegraphics[width=8cm]{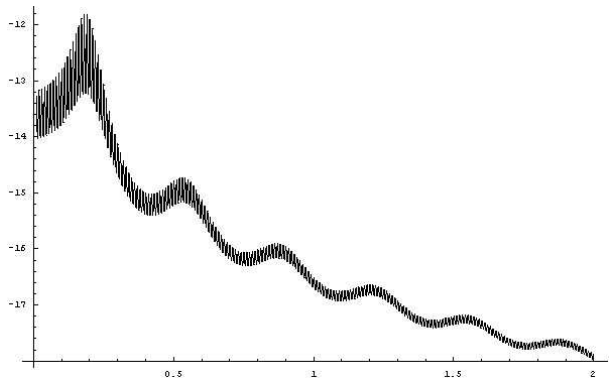} 
\caption{ Multi bump Flux Intensity in linear scale, normalized to
the visual magnitude for the same precessing Jet displayed in Figs.
3, 4. The profile shows the characteristic oscillatory damped decay
as that observed in GRB $030329$ and in  the $27$ August 1998 SGR.
The time scale is arbitrary: in the GRB $030329$ the unit
corresponds to about a  daily scale, while in SGR events the unit
is in minutes.}
\end{figure}
\begin{figure}\centering\includegraphics[width=8cm]{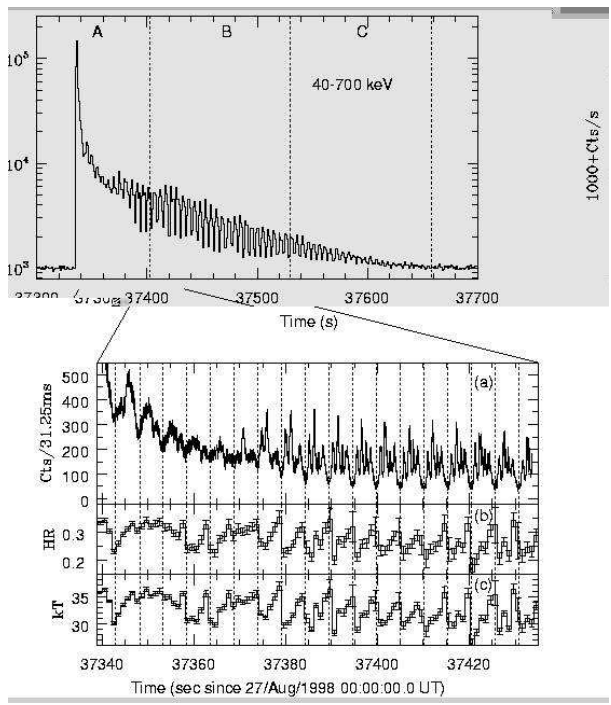} 
\caption[h]{The observed gamma profile of the intense SGR on $27$
August 1998. The burst decay seems to follow the damped oscillatory
behaviour shown above. }
\end{figure}
\begin{figure}\centering\includegraphics[width=8cm]{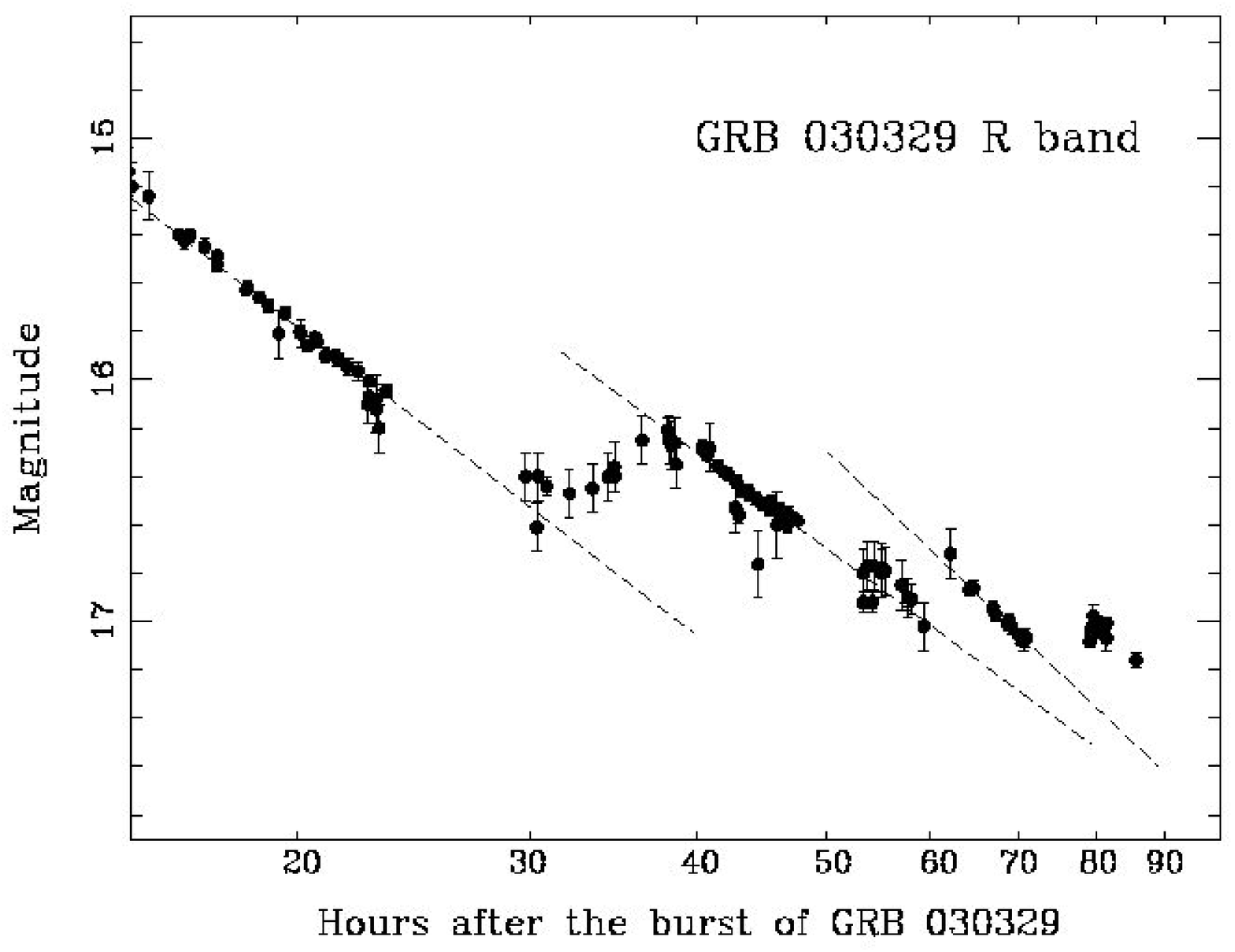} 
\caption {The multi bump behaviour or "re-brightening"
observed in GRB $030329$. Its puzzling imprint maybe described by a
precessing $\gamma$, $X$ and optical jet. }
\end{figure}
Fig. \ref{jet3D} displays a 3D pattern of the jet and Fig.
\ref{jet2D} its projection along the vertical axis in a 2D plane.
The combination of the different angular velocities determine the
multi-precession of the jet. Each component  includes the pulsar
jet spin angular velocity ($\omega_{psr}$) and its angle
$\theta_{psr}$, the nutation speed ($\omega_N$) and nutation angle
$\theta_{N}$ (due to possible inertial momentum anisotropies or
beam-accretion disk torques). A slower component due to the
companion of the binary system, $\omega_b$, (and the corresponding
angle $\theta_{b}$) will modulate the total jet precession. On
average, from Eq. 3 the $\gamma$ flux and the $X$, optical
afterglow are decaying as a power-law with time, $t^{-\alpha}$,
where $\alpha \simeq 1-2$. The spinning and precessing jet is
responsible for the wide range of GRBs and SGRs properties and of
their partial internal periodicity.
The $\gamma$ time evolution and spectra derived in this ideal model
may be compared successfully with observed GRB data (see Fig. 7,
8).


\section{Hard $\gamma$-X  Jet by Inverse Compton Scattering by electron pairs }
\label{sect:hard}
The $\gamma$ Jet is originated mainly by Inverse Compton
Scattering of GeVs electron pairs onto thermal photons (Fargion
1994; Fargion \& Salis 1995a, 1995c, 1998; Fargion 1999) in
nearly vacuum conditions. Therefore these electron pair are
boosted at Lorentz factor $\gamma_e \geq 2 \cdot 10^3$. Their
consequent Inverse Compton Scattering will induce a $\gamma$ Jet
whose beaming angle is $\Delta \theta \leq \frac{1}{\gamma}
\simeq 5\cdot10^{-4} rad \simeq 0.0285^o $ and a wider, less
collimated X, optical cone. These angles are compatible with the
beaming required to explain the transition from the SN to the GRB
power. In addition the electron pair Jet may generate  a
secondary synchrotron radiation component at radio energies, in
analogy with  BL Lac blazars where the hardest TeV $\gamma$
component is made by Inverse Compton Scattering and the
correlated X band emission is due to the synchrotron emission.
The inner jet is dominated by the harder photons while the
external cone contains softer $X$, optical and radio waves. The
jet aperture, according to the theory of relativity, would imply
$\theta \sim \frac{1}{\gamma_e}$, where $\gamma_e \simeq 10^3
\div 10^4$ (Fargion 1999; Fargion et al. 1996b). In a first
approximation the constraint on the gamma energy range
 is given by the Inverse Compton relation: $< \epsilon_\gamma
> \simeq \gamma_e^2 \, k T$ for $kT \simeq 10^{-3}-10^{-1}\, eV$
and $E_e \sim GeVs$ leading to characteristic X-$\gamma$ GRB
spectra.  $GeV$ electron pair are  likely to be related to primary
muon pair jets, able to cross dense stellar target (Fargion \&
Salis 1998). The consequent adimensional photon number rate,
 as a function of the angle
$\theta_1$ becomes
(Fargion 1999)
\begin{equation}
\label{eqn:eqn2} \frac{\left( \frac{dN_{1}}{dt_{1}\, d\theta
_{1}}\right) _{\theta _{1}(t)}}{ \left( \frac{dN_{1}}{dt_{1}\,
d\theta _{1}}\right) _{\theta _{1}=0}}\simeq \frac{1+\gamma
^{4}\, \theta _{1}^{4}(t)}{[1+\gamma ^{2}\, \theta
_{1}^{2}(t)]^{4}}\, \theta _{1}\approx \frac{1}{(\theta
_{1})^{3}} \;\;.\label{eq4}
\end{equation}
The total fluence at the minimal impact angle $\theta_{1 m}$
responsible for the average luminosity  is
$$
\frac{dN_{1}}{dt_{1}}(\theta _{1m})\simeq \int_{\theta
_{1m}}^{\infty }\frac{ 1+\gamma ^{4}\, \theta _{1}^{4}}{[1+\gamma
^{2}\, \theta _{1}^{2}]^{4}} \, \theta _{1}\, d\theta _{1}\simeq
\frac{1}{(\, \theta _{1m})^{2}}\;\;\;.
$$
These spectra fit GRBs observed ones (Fargion \& Salis 1995a,
1995c; Fargion et al. 1996b; Fargion 1999). Assuming a beam jet
intensity $I_1$ comparable with maximal SN luminosity, $I_1 \simeq
10^{45}\;erg\, s^{-1}$, and replacing this value in the above
adimensional equation  we find a maximal apparent GRB power for
beaming angles $10^{-3} \div 3\times 10^{-5}$, $P \simeq 4 \pi
I_1 \theta^{-2} \simeq 10^{52} \div 10^{55} erg \, s^{-1}$, just
within the observed values. We also assumed that the jet
intensity decays in time as the following power law
$$
  I_{jet} = I_1 \left(\frac{t}{t_0} \right)^{-\alpha} \simeq
  10^{45} \left(\frac{t}{3 \cdot 10^4 s} \right)^{-1} \; erg \,
  s^{-1}
$$
assuming that at a time scale of 1000 years it may reach  the
 observed intensity of known galactic micro-jets such as SS433: $I_{jet}
\simeq 10^{39}\;erg\, s^{-1}$. This offers a natural link between
the GRB and the SGR output powers. We used the model to evaluate
if the April precessing jet might hit us once again. It should be
noted that a steady angular velocity would imply an intensity
variability ($I \sim \theta^{-2} \sim t^{-2}$) corresponding to
some of the earliest afterglow decay law. These predictions have
been proposed a long time ago, (Fargion 1999). Similar
descriptions with more parameters and within a sharp time
evolution of the jet have been  also proposed by other authors
(Blackman, Yi, \& Field, 1996; Portegies Zwart et al. 1999).
\section{Precessing Radio Jet by Synchrotron Radiation}
\label{sect:prec}
As we have mentioned in the last section, the same GeV Jet of
electron pair may generate a secondary beamed synchrotron radiation
component at radio energies, in analogy to the behaviour of  BL Lac
blazars.
However the inner jet is dominated by harder photons while the
external cone contains softer $X$, optical and radio waves. Their
wide precessing  angle is the source of the radio bumps emitted on
a time scale of days, clearly observed in the GRB$980425$,
GRB$030329$ light curves. The peculiar and oscillating optical
variability of GRB$970508$ did show a re-brightening nearly  two
months later, and it did also show a remarkable multi-bump
variability in radio wavelength. For this reason  we are more
inclined to believe that this fluctuations were indeed related to
the Jet precession and not to any interstellar scintillation.
There is not any direct correlation between the $\gamma$ Jet made
up by Inverse Compton scattering and the Radio Jet because the
latter is dominated by the external magnetic field energy
density. There maybe a different beaming opening and a consequent
different time modulation respect to the inner $\gamma$ Jet.
However the present wide energy power emission between SN2002ap
and GRB$030329$ radio light curves makes probable a comparable
beaming angle: $\leq 10^{-3}- 10^{-4}$ radiant. The appearance of
re-brightening is not unique of GRBs and SGRs. It should also
occur in the X-Ray-Flare (XRF) event that have been recently
associated to GRBs. Indeed the recent XRF $030723$ (Butler et. al
2004 ) exhibit an X-Ray re-brightening; also the  Optical
Afterglow of GRB 021211 (Della Valle et al. 2003) did show such a
surprising increase in luminosity at very late times in the GRB
afterglow.

\section{X Ray precursor by Precessing Jet}
\label{sect:x ray}
\begin{figure}
\centering\includegraphics[width=10cm]{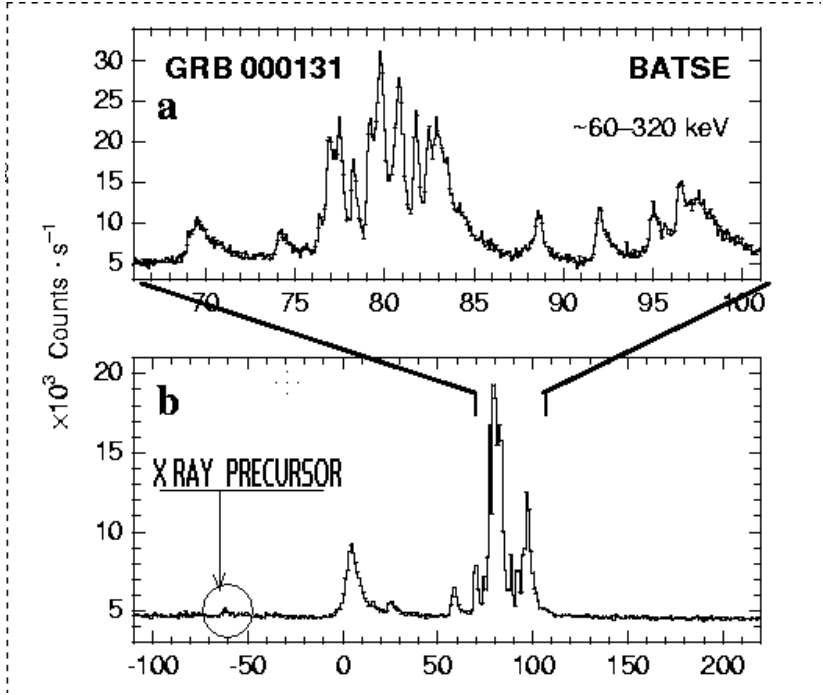}
\caption {X precursors in GRB000131, the most distant $\gamma$
burst observed so far ($z= 4.5$).}
\end{figure}

The thinner the jet, the larger the sample of events, the source
volume (large redshift) and the harder is the $\gamma$ Jet
observed. This explain why, despite the Hubble cosmic expansion and
 the time dilution, the most
variable and the most powerful  observed GRBs with the hardest
spectra are not the nearest ones but the most distant. Isotropic
explosions can not explain such a scenario. Indeed the extreme
$\gamma$ energy budget, requiring a comparable $\nu$ emission,
exceeds in isotropic models few or many solar masses,  even
assuming an (ideal) entire mass-energy conversion. To probe how
the Jet model may fit the GRB features, let us consider the  most
distant ($z= 4.5$) GRB observed so far - GRB$000131$ - and its $X$
ray precursor. This burst while being red-shifted and slowed down
by a factor $5.5$, shows a short time scale and a very fine
structure in disagreement with any fireball model, but well
compatible with a thin, fast spinning precessing $\gamma$ jet.
Moreover let us notice the presence of a weak $X$-ray precursor
pulse lasting 7 sec, 62 sec before the  main  $\gamma$ burst
trigger GRB$000131$ (Fargion 2001). Its arrival direction within
12 degree from the  GRB event is consistent only with the main
pulse (the probability for this coincidence to occur by chance is
below $3.6\cdot 10^{-3}$). Given the time clustering proximity
(one minute over a day GRB rate average),  the probability  to
occur by chance is below one thousandth. The overall probability
to observe this precursor by chance is below 3.4 over a million
making inseparable its association with the main GRB000131 event.
This weak burst signal corresponds to a power of more than a
million Supernovae and have left no trace or Optical/X transient
just a minute before the more energetic gamma event (peak power
$> billion $ Supernova). No isotropic GRB explosive progenitor
could survive such a disruptive  precursor  nor any
multi-exploding jet. Only a persistent, pre-existing precessing
Gamma Jet pointing twice near the observer direction could
naturally explain such a luminosity evolution. These X-ray
precursor are not  unique but are found in $3-6\%$ of all GRBs.
Similar X precursors occurred in SGRs event as the $1900 + 14$ on
29 August 1998.






\section{Conclusions: Neutrino-Muon Jets Progenitors }
\label{sect:conclusion}
We  believe that GRBs and SGRs  are persistent blazing flashes from
lighthouse-like, thin $\gamma$ Jets spinning  in multi-precessing
  modes (binary, precession, nutation). These GRBs Jets are
originated by NSs and/or BH in binary systems, or accretion disks
fuelled by infalling matter. Their relics (or they progenitors)
are nearly steady X-ray Pulsars whose fast blazing may be the
source of SGRs. The Jets are not  single explosive events in GRB,
but they are powered at maximal output during the long period of
a SN explosion. The power of the  beamed Jet is comparable to
that of a SN at the peak luminosity. The external $\gamma$ Jet is
originated by a series of processes linked one to the other.
First the jet may be originated in the SN and/or BH birth, and it
is probably due to a very collimated primary muon pair Jet at
TeVs-Pevs energies. These muons might propagate with negligible
absorption through the dense photon background produced by the SN
explosion, and they are nearly transparent to photon-photon
opacities. We speculate that such muon pair progenitors might be
secondaries of a ultra-high energy neutrino Jet, originated in
the interior of a new born NS or BH, because neutrinos are able
to escape from the dense matter envelopes obscuring the
Super-Nova volume (Gupta 2003; Fargion 2002). The high energy
relativistic muons (at tens of TeV-PeV) decay in flight in
electron pair where the baryon density is still negligible. The
Inverse Compton Scattering of such electronic pair with nearby
thermal photon is the final step to produce the observed hard $X$
- $\gamma$ Jet. The cost of this long chain of reactions  is a
poor energy conversion, but it has the advantage of being able to
explain the $\gamma$ escape from a very dense environment
"polluted" by matter and radiation. The relativistic morphology
of the Jet and its multi-precession geometry is the source of the
complex $X$-$\gamma$ spectra signature of GRBs and SGRs. The
inner Jet produces by relativistic Inverse Compton Scattering the
hardest and rarest beamed GeVs-MeVs photons (as the rare and long
$5000$ s life EGRET GRB$940217$ one). The external edges of the
Jet create softer and softer photons.

The complex variability of GRBs and SGRs is discussed and
compared. We find that the properties of both events are
successfully described  by a multi-precessing Jet whose angular
evolution is described by the Eq. 1 (see Fargion \& Salis 1995a,
1995b; Fargion 1999). Such a beamed Jet
 may also explain  the wide range of $X-\gamma$
signatures. Therefore the puzzle of GRBs is no longer in their
apparent huge luminosity, but in the mechanism able to originate
such an extreme jet collimation and its precession.



\label{lastpage}



\end{document}